\documentclass[preprint,12pt,3p,twocolumn,letterpaper]{elsarticle}

\journal {Physics Letters B}
\begin{document} 

\begin{frontmatter}

\title{Pair-vibrational states in the presence of neutron-proton pairing}

\author{R.~R.~Chasman\fnref{fn1}}
\address{Physics Division, Argonne National Laboratory, Argonne Illinois 60439-4843}

\author{P.~Van~Isacker\fnref{fn2}}
\address{Grand Acc\'el\'erateur National d'Ions Lourds, CEA/DSM-CNRS/IN2P3\\
B.P. 55027, F-14076 Caen Cedex 5, France}
\fntext[fn1]{chasman@anl.gov, ~Corresponding Author}
\fntext[fn2]{isacker@ganil.fr}

\begin{abstract}
Pair vibrations are studied for a Hamiltonian with neutron-neutron,
proton-proton and neutron-proton pairing.  The spectrum is found to be
rich in strongly correlated, low-lying excited states.  Changing the
ratio of diagonal to off-diagonal pairing matrix elements is found to
have a large impact on the excited-state spectrum.  The variational
configuration interaction (VCI) method, used to calculate the
excitation spectrum, is found to be in very good agreement with exact
solutions for systems with large degeneracies having equal $T=0$ and
$T=1$ pairing strengths.
\end{abstract} 

\begin{keyword}
N-P Pairing \sep Excited States \sep Variational Configuration Interaction Method
\PACS 21.10.Dr, 21.30.Fe, 21.60.Fw, 21.60.-n, 31.15.-p 
\end{keyword}

\end{frontmatter}

\section{Introduction}
The collective features of many quantum systems are driven by the
tendency of the constituent particles to form pairs which translates
into the existence of a strong attractive pairing interaction.  In
atomic nuclei the correlations between pairs of nucleons give rise to
pair-vibrational states which occur when nucleons are collectively
excited across a gap in the single-particle spectrum~\cite{BM75}.  The
conditions for the existence of such states are level degeneracies
just above and below the gap and an appropriate value for the pairing
interaction strength.  Given these conditions, there exist excited
(pair-vibrational) states with energies substantially less than the
energy required to promote a pair of nucleons across the gap.  In such
cases the correlation energy in the excited state is comparable to or
larger than it is in the ground state.

The problem of pair vibrations has been previously studied for a
Hamiltonian with like-particle pairing~\cite{BM75,Hogaasen61}.
However in systems such as nuclei with two different kinds of
particles, the pairing interaction has both isoscalar ($T=0$) and
isovector ($T=1$) components.  The interaction between identical
nucleons is of pure isovector character while the neutron-proton
interaction has both an isoscalar and an isovector component.  Since
there is no {\it a priori} reason for the isoscalar pairing to
vanish---empirically it is found to be of the same order as isovector
pairing---both modes are present in nuclei, rendering the problem of
pair vibrations in nuclei considerably more complex but also richer.

In this Letter we investigate the nature of pair-vibrational states in
the presence of isoscalar and isovector pairing.  Using a variational
configuration-interaction (VCI)~\cite{Chasman05} method, we solve a
many-body problem for neutrons and protons distributed over levels
interacting through the two types of pairing with adjustable
strengths.  This method is tested by comparing to solvable cases and
is found to be excellent for the lowest eigenstates.  We then present
results for more realistic choices of the strengths and parameters of
the model Hamiltonian, and study the influence of these modifications
on the character of the pair vibrations.

\section{Model, Calculations and Results}
Our Hamiltonian is~\cite{Goodman79,Chasman03} 
\begin{eqnarray}
\lefteqn{H}
&\;=&\!\!\sum_{i>0}
\epsilon_i\left(
a_i^\dag a_i\!+\!a_{-i}^\dag a_{-i}\!+\!
b_i^\dag b_i\!+\!b_{-i}^\dag b_{-i}\right)
\nonumber\\
&-&\!\!\sum_{i,j} G_{i,j}^{T=1}
\left[A_i^\dag A_j+B_i^\dag B_j+C_i^\dag C_j\right]
\nonumber\\
&-&\!\!\sum_{i,j} G_{i,j}^{T=0}
\left[(M_i^\dag M_j+N_i^\dag N_j)
\delta_{\Omega_i\Omega_j}\right]
\nonumber\\
&-&\!\!\sum_{i,j} G_{i,j}^{T=0}
\left[D_i^\dag D_j\right],
\label{e_ham}
\end{eqnarray}
where, in the spherical limit, $i$ denotes a single-particle state
with spin projection $j_z=\pm \Omega_i$ and energy $\epsilon_i$.  In
the deformed limit, $i$ denotes a Nilsson orbital whose projection on
the nuclear symmetry axis is $\pm\Omega_i$.  In either case, each
level $i$ accommodates at most two neutrons and two protons.  The
operators $a_i^\dag$ and $b_i^\dag$ create a neutron (n) and a proton
(p), respectively.  Furthermore, the n-n and p-p pair creation
operators are $A_i^\dag=(a_i^\dag a_{-i}^\dag)$ and
$B{_i}^\dag=(b_i^\dag b_{-i}^\dag)$.  The $T=1$ n-p pair creation
operator is $C{_i}^\dag=\frac{1}{\sqrt 2}(a_i^\dag
b_{-i}^\dag+a_{-i}^\dag b_i^\dag)$ and the $T=0$ n-p pair creation
operator is $ D{_i}^\dag=\frac{i}{\sqrt 2}(a_i^\dag
b_{-i}^\dag-a_{-i}^\dag b_i^\dag)$.  Also, $M_i^\dag=(a_i^\dag
b_i^\dag)$ and $N_i^\dag=(a_{-i}^\dag b_{-i}^\dag)$.  We choose the
proton wave function with $j_z=1/2$ as the negative of the equivalent
neutron wave function.

In the model system that we use to elucidate the properties of pair
vibrations, we take eight levels at $\epsilon=0$~MeV and eight levels
at $\epsilon=1$~MeV. Each levels has an angular momentum $J=1/2$ and
accomodates two neutrons and two protons.  Because all levels have
$J=1/2$, the $T=0$ $M_i^\dag$ and $N_i^\dag$ modes play an important
role in determining correlations in the wave functions and are in fact
equivalent to the $A_i^\dag$ and $B_i^\dag$ $T=1$ modes.  We consider
16 neutrons and 16 protons here so that for zero pairing strengths the
ground state has all levels filled up to the gap and the lowest pair
excitations are at 2~MeV.  We carry out calculations for several
(off-diagonal) interaction strengths $G^{T=1}=0.05$, 0.10, 0.15, 0.20
and 0.30~MeV.  We consider three cases.

{\it Case 1.}  We set $G^{T=0}=G^{T=1}$ and take equal diagonal and
off-diagonal matrix elements; diagonal matrix elements refer to $i=j$
in Eq.~(\ref{e_ham}) while off-diagonal ones involve $i\neq j$
interactions.  It is known from the work of Flowers and
Szpikowski~\cite{Flowers64} that a system of neutrons and protons,
interacting through pairing with equal $T=0$ and $T=1$ strengths and
occupying degenerate orbits, is analytically solvable because of an
underlying dynamical symmetry associated with the SU(4) algebra which
occurs as a subalgebra of the total n-p pairing algebra SO(8).  If we
group the single-particle levels into two sets at different energies,
we obtain a two-level SO(8) problem which is no longer analytically
but still numerically solvable through the diagonalization of matrices
of modest size~\cite{Pang69,Dussel86}.  This is true for the lowest
eigenstates of the pairing Hamiltonian which are of low seniority
(provided the pairing force is attractive) and for which the necessary
coupling coefficients [for ${\rm SO(8)}\supset{\rm SU}(4)$ and ${\rm
  SU}(4)\supset{\rm SU}_S(2)\otimes{\rm SU}_T(2)$] are
known~\cite{Pang69}.  The current problem of eight-plus-eight $J=1/2$
levels is equivalent to a two-level SO(8) description in which each
level has a spatial degeneracy of eight, {\it i.e.}, each level can
accommodate 16 neutrons and 16 protons.  Exact energies can also be
obtained with the Richardson-Gaudin method for the higher-rank algebra
SO(8) for non-degenerate levels with equal $T=0$ and $T=1$ pairing
strengths~\cite{Lerma07}.  These exact solutions provide a very
valuable test for the approximate solutions described below.

{\it Case 2.}  We again set $G^{T=0}=G^{T=1}$ but diagonal matrix
elements are now 2.4 times the off-diagonal ones. Our motivation is
that the increased value of the diagonal matrix elements explains the
Wigner energy anomaly~\cite{Chasman03} and also the discrepancy
between `observed' single-particle gaps and those obtained from
Woods-Saxon potentials in $N=Z$ nuclei~\cite{Chasman07}.  In the SO(8)
model of Refs.~\cite{Pang69,Dussel86} it is also possible to take
unequal diagonal and off-diagonal pairing strengths, if only two
levels are considered.  However, in the current application we have 16
levels.  In this case, the unequal diagonal and off-diagonal pairing
strengths suggested by physical arguments, no longer permit the same
solution technique as in Refs.~\cite{Pang69,Dussel86}.

{\it Cases 3 and~3'.}  We set diagonal matrix elements to be 2.4 times
the off-diagonal ones and we change the relative strengths of $T=0$
and $T=1$ pairing, {\it i.e.}, in case~3 we set $G^{T=0}=0.9\times G^{T=1}$
and in case~3' we interchange the pairing strengths.  Case~3 suggests
the sorts of differences to be expected in heavier nuclei where
odd-odd $N=Z$ nuclei have a $0^+$ ground and a $1^+$ excited state.
Case~3' suggests light nuclei where the $0^+$ state is not the ground
state in odd-odd $N=Z$ nuclei.  The energies are the same in cases~3
and 3' but the $(J,T)$ labels of the states are interchanged.

For the even-even systems considered here,
we use a variational wave function of the form
\begin{equation}
\Theta_i
={\cal P}\prod^k\psi^\dag_{i,k}|0\rangle,
\end{equation}
where $|0\rangle$ is the physical vacuum
and $\psi^\dag_{i,k}$ is a creation operator of the form
\begin{eqnarray}
\psi^\dag_{i,k}&=&
\left[1+U_i(1,k)A_k^\dag+U_i(2,k)B_k^\dag\right.
\nonumber\\
&&
+U_i(3,k)C_k^\dag+U_i(4,k)D_k^\dag
\nonumber\\
&&+U_i(5,k)M_k^\dag +U_i(6,k)N_k^\dag
\nonumber\\
&&\left.+U_i(7,k)W_k^\dag\right],
\end{eqnarray}
with
$W_k^\dag=A_k^\dag B_k^\dag$.
This is an extended version of the
variational wavefunction used in previous
studies~\cite{Chasman05,Chasman03,Chasman07} by virtue of the addition
of the $M_k^\dag$ and $N_k^\dag$ terms which are needed because all
$j_z$ values are the same and these modes are collective.  The
operator ${\cal P}$ projects definite neutron number, proton number,
number parity of $T=0$ n-p pairs~\cite{Chasman03} and now also
$J_{z}$.  We project before carrying out the variational procedure.

The variational trial wave function, $\Xi^{n+1}_m$,
is~\cite{Chasman05}
\begin{equation}
\Xi^{n+1}_m=\Phi^n_m+\Theta_{n+1},
\end{equation}
where $\Phi^n_m$, the starting wave function, is
\begin{equation}
\Phi^n_m=\sum_{i=1}^n t^n_{i,m}\Theta_i, 
\end{equation}
with $n$ the number of VCI basis states, and $m$ the specific state
(ground or excited state) that we are approximating.  All $\Theta_i$
have exactly the same structure differing only in the numerical values
of the amplitudes $U_i(j,k)$.
Each of the fully projected states
$\Theta_i$ consists of $1.108\times10^{12}$ Slater determinants.

For case~1, we have exact results for comparison.  In
Table~\ref{tab:case1} we list the exact energies for case~1. The
exact solutions provide valuable tests for future approximate
methods. Our approximation is in extremely good agreement with the
exact results over the entire range of interaction strengths. The
ground state and first four excited states for even isospin $T$ are
obtained accurately for 150--200 VCI basis states.  For odd isospin
$T$, we consider only the lowest solution as it is the only one in the
energy range of the four low-lying $T$-even excited states.  The
even-$T$ ground state and the lowest $T$-odd state are approximated to
a few keV accuracy. This seems to be a general feature of our
approximate method - the lowest eigenstate for a given set of quantum
numbers can be obtained more accurately than excited states having
those quantum numbers. The first three even-$T$ excited states are
accurate to 10--20~keV and the fourth state to 40--50~keV for the
largest values of the interaction strength and somewhat better for
weaker interaction strengths. The agreement in energy is quite
satisfactory and can be further improved by increasing the number of
VCI configurations.

\begin{table*}
\centering
\caption{Exact energies (in MeV) for case~1.}
\label{tab:case1}
\begin{tabular}{cccccccc}
&&&&&&&\\
\hline
 & $G^{T=0,1}$ & $E^{T={\rm e}}_0$ & $E^{T={\rm o}}_0$
 & $E^{T={\rm e}}_1$ & $E^{T={\rm e}}_2$ & $E^{T={\rm e}}_3$ & $E^{T={\rm e}}_4$   \\
\hline
 & 0.05 &$-3.0146$&$-1.5206$&$-1.5206$&$-1.5206$&$-1.5206$&$-1.4482$\\
 & 0.10 &$-7.7112$&$-6.6071$&$-6.6071$&$-6.6071$&$-6.6071$&$-6.4395$\\
 & 0.15 &$-14.401$&$-13.299$&$-13.299$&$-13.299$&$-13.299$&$-12.526$\\
 & 0.20 &$-22.181$&$-20.852$&$-20.852$&$-20.852$&$-20.852$&$-19.327$\\
 & 0.30 &$-38.780$&$-36.902$&$-36.902$&$-36.902$&$-36.902$&$-34.189$\\
\hline
\multicolumn{8}{l}{The superscripts $T={\rm e}$ and $T={\rm o}$
denote $T$-even and $T$-odd states.}
\end{tabular}
\end{table*}

\begin{figure*}
\centering
\includegraphics[width=15cm]{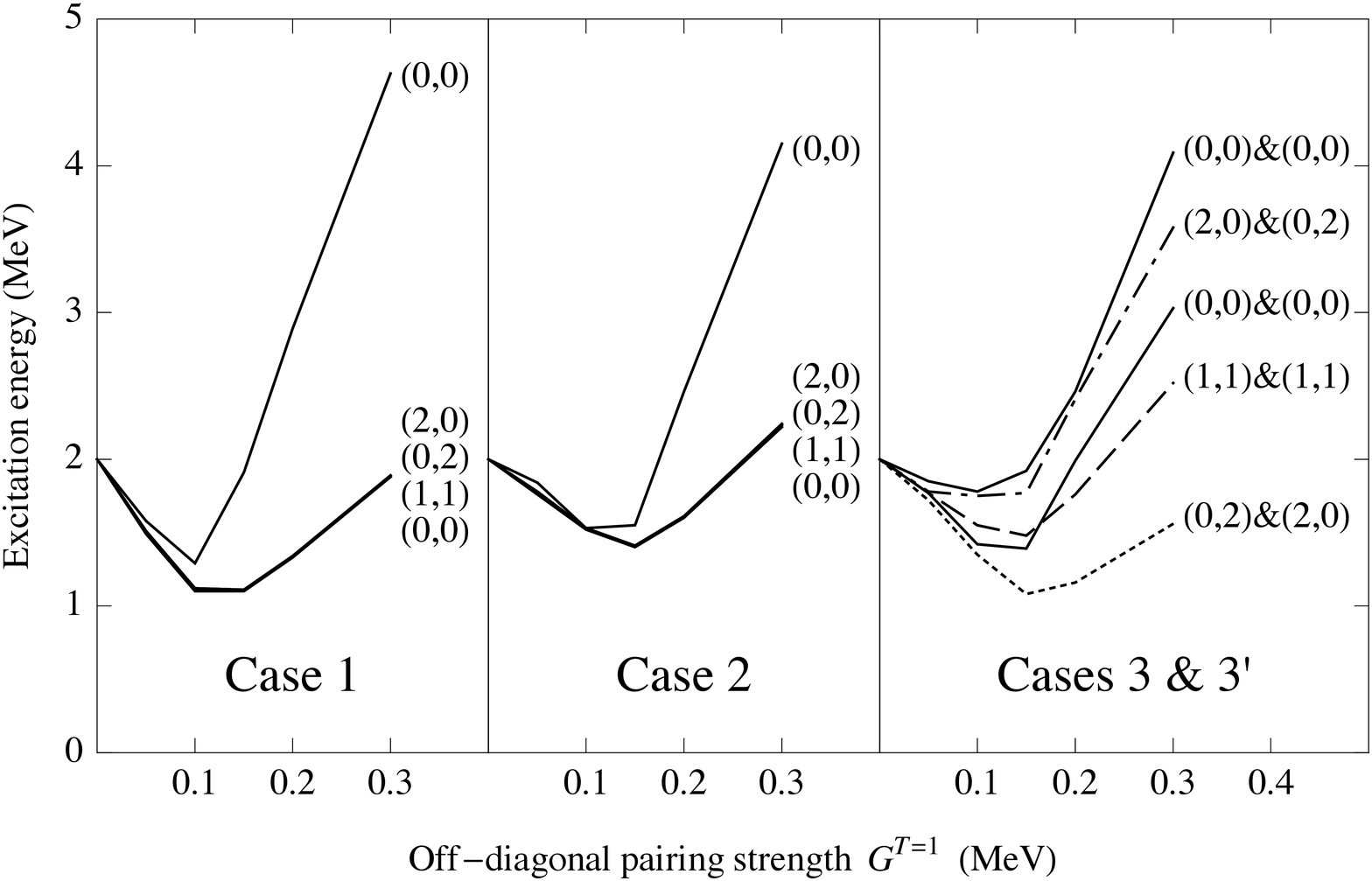}
\caption{Excitation spectra as a function of the off-diagonal pairing strength $G^{T=1}$.
The leftmost spectrum is obtained for the parameters of case~1, described in the text,
the center spectrum is for case~2,
and the rightmost for cases~3 and~3'.
Levels are labelled with the quantum numbers $(J,T)$.}
\label{fig:nuvib}
\end{figure*}

In Fig.~\ref{fig:nuvib} we show the spectra calculated for the four
cases~1-3'.  For cases~1 and 2 the four lowest excited states are
degenerate, one has odd isospin and the other three have even isospin.
The $(J,T)$ assignments of these four states are (0,0), (1,1), (2,0),
and (0,2).  The fifth excited state is (0,0).  Although the
excited-state degeneracy pattern is the same for cases~1 and 2, the
excitation energies of the states in the quadruplet differ
substantially.  This has important implications for the relation
between relative $T=0$ and $T=1$ pairing strengths and moments of
inertia.  The only difference between the two cases is in the ratio of
off-diagonal to diagonal matrix elements.  Yet the moment of inertia,
as determined by the energy difference of the first $2^+$ state and
the $0^+$ ground state, changes by 30--40\% between the two cases.

In case~3 a 10\% reduction of the $T=0$ strength from the values in
case~2 leads to large changes in the excited-state spectrum and in the
moment of inertia. As there are no exact solutions for guidance in
this case, we can get some sense of the quality of the VCI method for
excited states by comparing the energy of the $J_z=0$ member of the
J=2 multiplet with the $J_z=2$ member of the same multiplet. The
$J_z=2$ member of the multiplet is the lowest $J_z=2$ state and can be
calculated more accurately. The $J_z=2$ state is calculated to be
roughly 10 keV below the states labelled (2,0), giving us further
confidence in the VCI method for excited states. We have calculated
the lowest $J_z=2$ state for both cases~3 and 3'. 

The similarity of the energies of the lowest $J=2^+$ in case~3' and
case~1 should make one cautious about inferring anything about
relative $T=0$ and $T=1$ interaction strengths from moments of inertia
as the $2^+$ excitation energies depend strongly on both the relative
interaction strengths and diagonal matrix element strengths.  The
effects of large diagonal matrix elements, when properly taken into
account by including $W^\dag$ terms, should persist to higher spins.

\begin{table*}
\centering
\caption{Off-diagonal ground-state correlation energies (in MeV).}
\label{tab:correl}
\begin{tabular}{ccccccccc}
&&&&&&&&\\
\hline
 && $A^\dag A$ & $B^\dag B$ & $C^\dag C$ & $D^\dag D$ & $M^\dag M$ & $N^\dag N$ &  \\
\hline
 & Case 1  & 1.21 & 1.21 & 1.20 & 1.20 & 1.21 & 1.21 \\
 & Case 2  & 0.92 & 0.92 & 0.91 & 0.91 & 0.92 & 0.92 \\
 & Case 3  & 1.07 & 1.07 & 1.06 & 0.61 & 0.62 & 0.62 \\
 & Case 3' & 0.62 & 0.62 & 0.61 & 1.06 & 1.07 & 1.07 \\
 \hline
\end{tabular}
\end{table*}
\begin{table*}
\centering
\caption{Off-diagonal correlation sums (dimensionless).}
\label{tab:totcor}
\begin{tabular}{cccccccccc}
&&&&&&&&&\\
\hline
&& G.S. & $1*$ & $2*$ & $3*$ & $4*$ & $5*$ & $6*$ & $7*$ \\
\hline
& Case 1e & 1.05 & 1.01 & 1.01 & 1.01 & 0.91 & 0.95 & 0.95 & 0.95 \\
& Case 1o & 1.01 & 0.95 & 0.95 & 0.95 & --- & --- & --- & --- \\
& Case 2e & 1.01 & 0.98 & 0.98 & 0.97 & 0.85 & 0.85 & 0.92 & 0.91 \\
& Case 2o & 0.98 & 0.92 & 0.92 & 0.92 & --- & --- & --- & --- \\
& Case 3e & 0.98 & 0.92 & 0.91 & 0.90 & 0.83 & 0.91 & 0.85 & 0.83 \\
& Case 3o & 0.93 & 0.89 & 0.86 & 0.85 & --- & --- & --- & --- \\
& Case 4e & 0.95 & 0.82 & 0.82 & 0.72 & 0.72 & --- & --- & --- \\
\hline
\multicolumn{10}{l}{An asterisk denotes an excited state.}
\end{tabular}
\end{table*}
Not only are excitation spectra affected by these differences,
ground-state correlation energies also change noticeably in going from
case~1 to case~2 and differ substantially in cases~3 and 3'.  In
Table~\ref{tab:correl} we compare the ground-state off-diagonal
correlation energies in the three cases for an off-diagonal strength
$G^{T=1}=0.1$~MeV.  For example, a value of 1.21~MeV for $A^\dag A$
means $\langle\sum_{i\neq j}G^{T=1}_{i,j}A_i^\dag
A_j\rangle=1.21$~MeV.  It is the off-diagonal correlation energy that
measures the collectivity of a state. Diagonal correlation energies
are typically large, whether or not a state is collective.  For a
Slater determinant configuration, there is no collectivity and the
off-diagonal correlation energy vanishes.  In our model system, the
diagonal correlation energy is $24\times(G^{T=1}_{i,i}+G^{T=0}_{i,i})$ for
the Slater determinant configuration. A 10\% reduction in $T=0$ matrix
elements reduces the off-diagonal $T=0$ correlation energy in the
ground state by roughly 40\%.  Small changes in the relative strengths
are greatly amplified in the wavefunctions.  If the wave functions
were exact, the values for the first three entries in each line would
be identical, as would be the last three.

In Table~\ref{tab:totcor} we compare the sum of all off-diagonal
correlation energies for the first eight $T$-even approximate states
and the first four $T$-odd states for $G^{T=1}=0.3$ for all three
cases.  In addition, we have included results for a system with only
like-particle pairing, labeled case~4.  Although we have not carried
out any minimizations of the $T$-even states 6-8 or for
$T$-odd states 2-4, the energies are in moderately good agreement with
the exact results calculated for case~1.  We choose $G^{T=1}=0.3$
because this is the largest value for which we have carried out
calculations.  Collectivity persists in all of the excited states. In
a system with only n-n and p-p pairing, the ground state off-diagonal
correlation energy is $2G\times P(L-P)$ in the degenerate limit, where
$P(=8)$ is the number of like-particle pairs and $L(=16)$ is the
number of levels; so we have divided all correlation energy sums by
$128G^{T=1}$.  The correlation energy of excited states drops off
somewhat faster in case~4 than in the cases with n-p pairing.  This
suggests that many excited states will show collective features
in nuclides near the $N=Z$ line.

\section{Summary}
In summary, we have applied the VCI method to the pair-vibrational
excited states of a Hamiltonian with $T=0$ and $T=1$ pairing.  We find
that there are at least five excited states with energies below the
two-nucleon excitation energy ({\i.e.}, pair vibrations) over much of
the range of vibrational interaction strengths.  This contrasts with
systems having only n-n and p-p pairing where only two low-lying
pair-vibration states~\cite{Hogaasen61} exist.  The large off-diagonal
correlation energies (indicative of collectivity) of these excited
states suggests that many collective states should be seen in nuclei
near the $N=Z$ line.  We find that the moment of inertia is very
sensitive to the ratio of diagonal to off-diagonal matrix elements, as
well as expectedly sensitive to the ratio of $T=0$ to $T=1$
interaction strengths.  Comparing VCI results with exact calculations
shows that the VCI method gives accurate energies for both ground and
excited states, in systems with large single-particle energy level
degeneracies.  In a previous study~\cite{Chasman05} the method was
successfully applied to systems with non-degenerate levels.  The VCI
method is quite general in that it does not impose any restrictions on
energies or matrix elements.

\section*{Acknowledgements}
We thank David Jenkins and Iain Moore for inviting us to participate
in an ECT* workshop on $N=Z$ nuclei where this collaboration started.
We thank A. Afanasjev for stimulating our interest in $T=0$ pairing.
We thank H. Esbensen for a useful discussion on this work.  Most of
the calculations were carried out on the Jazz computer array at
Argonne National Laboratory and on the Particle Physics Linux Farm at
the Weizmann Institute. R. C. thanks the Minerva Foundation for
supporting his stay at the Dept. of Physics of the Weizmann Institute
where much of this work was done.  The work of R.C. is supported by
the U.S. Department of Energy, Office of Nuclear Physics, Contract No.
DE-AC02-06CH11357; P.V.I is in part supported by the Agence Nationale
de Recherche, France, under contract nr ANR-07-BLAN-0256-03.

\end{document}